\documentclass[letterpaper, twocolumn, 9 pt, journal]{IEEEtran}  

\IEEEoverridecommandlockouts                           	

\usepackage[latin1]{inputenc}																						%
\usepackage[english]{babel}                                             %
\usepackage[T1]{fontenc}																								%
\usepackage{amsmath, amssymb, amsfonts}																	%
\usepackage{amstext}																										%
\usepackage{type1cm}																									  %
\usepackage{type1ec}																										%
\usepackage{dsfont}																										  %
\usepackage{booktabs}																									  %
\usepackage{array}																											%
\usepackage{enumerate}																									%
\usepackage[table]{xcolor}																							%
\usepackage{makeidx}																										%
\usepackage{mathtools}																									%
\usepackage{fancyhdr}																									  %
\usepackage{bm}																												  %
\usepackage{lastpage}																									  %
\usepackage{multicol}																										%
\usepackage{ifthen}																										  %
\usepackage{ifpdf}																											%
\usepackage{framed}																										  %
\usepackage[amsmath,thmmarks,framed,hyperref]{ntheorem}								  %
\DeclareMathAlphabet{\mathpzc}{OT1}{pzc}{m}{it}													%
\usepackage{psfrag} 																										%
\usepackage{float} 																											%
\usepackage{cite} 																											%
\usepackage[dvips]{graphicx} 																						%
\usepackage{subfloat} 																									%
\usepackage{stackrel} 																									%
\usepackage{graphicx} 																									%
\usepackage{textcomp} 																									%
\usepackage{color} 																											%
\usepackage{bbm} 																											  %
\usepackage{lettrine}																										%
\usepackage{dsfont} 																										%
\usepackage{multirow}																									  %
\usepackage[section,verbose]{placeins} 																	%
\usepackage{url}																												%
\usepackage{verbatim}																									  %
\usepackage{moreverb}																									  %
\usepackage{placeins}																										%
\usepackage{tikz}																											  %
\usetikzlibrary{arrows,automata,matrix,fit,positioning,calc}            %
\usepackage{pgfplots}																									  %

\newcommand{\I}{\boldsymbol{I}}

\newcommand{\x}{\boldsymbol{x}}

\usepackage{cite}
\usepackage{pgfplots}
\usetikzlibrary{arrows,automata,matrix,fit,positioning,calc}

\newcounter{thmCounter}
\newtheorem{definition}[thmCounter]{\bfseries Definition}
\newtheorem{theorem}[thmCounter]{\bfseries Theorem}

\newtheorem{assumption}[thmCounter]{\bfseries Assumption}

\newtheorem{lemma}[thmCounter]{\bfseries Lemma}
\newtheorem{remark}[thmCounter]{\bfseries Remark}

\usepackage{lipsum}

\usepackage{xcolor}

\usepackage{graphicx}

\def\x{\bm{x}}\def\y{\bm{y}}\def\w{\bm{w}}\def\v{\bm{v}}\def\0{\bm{0}}\def\bu{\bm{u}}\def\e{\bm{e}}

\def\beps{\boldsymbol{\epsilon}}
\def\boldeta{\boldsymbol{\eta}}
\def\I{\mathbf{I}}

\allowdisplaybreaks

\title{\LARGE \bf
Trade-Offs in Stochastic Event-Triggered Control
}

\author{Burak Demirel, Alex S. Leong, Vijay Gupta and Daniel E. Quevedo
\thanks{B. Demirel, A. S. Leong and D. E. Quevedo are with the Chair of Automatic Control (EIM-E), Paderborn University, Warburger Stra\ss e 100, 33098, Paderborn, Germany (e-mail: burak.demirel@protonmail.com, alex.leong@upb.de, dquevedo@ieee.org).
        }
\thanks{V. Gupta is with the Department of Electrical Engineering, University of Notre Dame, South Bend, Indiana, 46556, USA (e-mail: vgupta2@nd.edu).
		}
}

\begin{document}

\maketitle
\thispagestyle{empty}
\pagestyle{empty}

\begin{abstract}
This paper studies the optimal output-feedback control of a linear time-invariant system where a stochastic event-based scheduler triggers the communication between the sensor and the controller. The primary goal of the use of this type of scheduling strategy is to provide significant reductions in the usage of the sensor-to-controller communication and, in turn, improve energy expenditure in the network. In this paper, we aim to design an admissible control policy, which is a function of the observed output, to minimize a quadratic cost function while employing a stochastic event-triggered scheduler that preserves the Gaussian property of the plant state and the estimation error. For the infinite horizon case, we present analytical expressions that quantify the trade-off between the communication cost and  control performance of such event-triggered control systems. This  trade-off is confirmed quantitatively via numerical examples.
\end{abstract}


\section{Introduction}\label{sec:introduction}
Over the past decade, distributed control and estimation over networks have been a major trend. Thanks to the forthcoming revolution of the Internet-of-Things (IoT) and resulting interconnectedness of smart technologies, the importance of decision making over communication networks grows ever larger in our modern society. These technological advances, however, bring new challenges regarding how to use the limited computation, communication, and energy resources efficiently. Consequently,  event- and self-triggered algorithms have appeared as an alternative to traditional time-triggered algorithms in both estimation and control; see, e.g.,~\cite{HJT:12}.

A vast majority of the research in this area has mainly focused on proving the stability of the proposed control schemes, and demonstrating the effectiveness of such control systems, as compared to periodically sampled ones, through numerical simulations. However, an important stream of work in such schemes is \emph{analytically} characterizing the trade-off between the control performance and communication rate achieved via these algorithms. Early works on event-triggered control, such as~\cite{AsB:02, Rab:06, ImB:06, HJC:08}, provided performance expressions but only for scalar systems. The authors of~\cite{MeC:12} later extended the work of~\cite{AsB:02} to a class of second-order systems. The work in~\cite{MGA:17} studies state estimation for multiple plants across a shared communication network, and quantified communication and estimation performance. Recently, the authors of~\cite{GoM:17} investigated the minimum-variance event-triggered output-feedback control problem; cf.~\cite{AsB:02}. They established a separation between the control strategy and the scheduling decision, and they also showed that scheduling decisions are determined by solving an optimal stopping problem. Our initial work in~\cite{DGQ+:17b} considered a certain structure of controllers such as dead-beat controllers, and analytical expressions for the control performance and communication rate were obtained. Differing from~\cite{DGQ+:17b}, in the current work, we will focus on designing optimal controllers by establishing a separation between the controller and the scheduler.


Optimal event-triggered controller design requires the joint design of an optimal control law and an optimal event-based scheduler. The associated optimization problem becomes notoriously difficult~\cite{MoH:13} since, in general, the controller and the scheduler have different information. A vast majority of work in the literature focuses on the design of optimal feedback control laws for a predefined scheduling rule. It is important to note that designing an optimal control law might be very complicated even if one considers a fixed event-triggering policy. For instance, our recent work~\cite{DGQ+:17} shows that the optimal control problem, where a threshold-based event-triggered mechanism is used to decide if there is a need for transmission of new control actions based on knowledge of the plant state, leads to a non-convex optimization problem.

The selection of the event-triggering mechanism is essential for the computation of the control performance. As noted in~\cite{DGQ+:17b}, even in the case of Gauss-Markov plant models, due to the use of a (deterministic) threshold-based triggering mechanism, the plant state becomes a truncated Gaussian variable. As a result, computation of the control performance becomes challenging since it requires calculating the covariance of the plant state via numerical methods. One way to tackle this problem consists in employing a stochastic triggering mechanism, which preserves the Gaussianity of the plant state, as proposed in~\cite{HMW+:15, WRH+:16, WMS+:16}. Our initial work in~\cite{DLQ:17} used a deadbeat controller and a stochastic scheduler, which is similar to the ones mentioned above, to quantify the trade-off between the communication and the control cost for scalar systems. Similarly, the authors of~\cite{BAA:17} proposed an event-triggered control scheme that works under stochastic triggering rules. They also derived a control policy that always outperforms a periodic one. Differing from~\cite{BAA:17}, in this work, we will focus on solving the optimal output event-triggered control problem. 

{\it Contributions.}
In this paper, we consider optimal output-feedback control of a linear time-invariant system where a stochastic event-based triggering algorithm dictates the communication between the sensor and the controller. The proposed scheduler decides at each time step whether or not to transmit new state estimates from the sensor to the controller based on state estimation errors. 
%
The main contributions of this manuscript are as follows:  
\begin{itemize}
  \item[1)] We develop a framework for quantifying the closed-loop control performance and the communication rate in the channel between the sensor and the controller.
  \item[2)] We confirm that the certainty-equivalent controller is optimal under the scheduling rule based on estimation errors.  Our previous work~\cite{DGQ+:17b} used a transmission strategy based on the plant state, and employed a sequence of deadbeat control actions to establish a resetting property, but this was not optimal since the separation principle between control and scheduling does not hold. 
  \item[3)] We derive analytical expressions for the (average) communication rate and control performance. Our analysis relies on a Markov chain characterization of the evolution of the state prediction error (cf.~\cite{DGQ+:17b}) where the states of this Markov chain describe the time elapsed since the last transmission.
  \item[4)] Due to the use of the stochastic triggering rule, we can compute the conditional covariance of the comparison error (i.e., the difference between the state estimation error at the sensor and the state estimation error at the controller) in a closed-form. Consequently, it becomes almost effortless to compute the closed-loop control performance; cf.~\cite{DGQ+:17b}. 
\end{itemize}

{\it Outline.}
The remainder of the paper is organized as follows: Section~\ref{sec:problem_formulation} describes the system architecture and formulates an optimal event-triggered control problem. In Section~\ref{sec:main_results}, a control policy which minimizes a quadratic cost function under an event-triggered transmission constraint, is derived. This section also presents analytic expressions of the communication rate and the control performance for the infinite horizon problem. An illustrative example is presented to demonstrate the trade-off between communication and control performance in Section~\ref{sec:numerical_example}, while Section~\ref{sec:conclusions} finalizes the paper with concluding remarks. The Appendix provides detailed proofs of the main results.

%

\section{Problem Formulation} \label{sec:problem_formulation}

\begin{figure}[!t]\centering
	\includegraphics[scale=0.8]{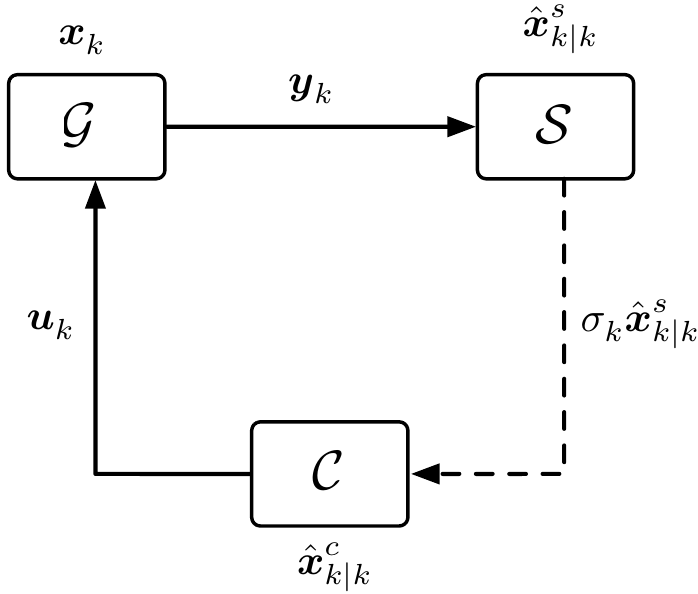}
	\caption{Block diagram of a feedback control system with plant $\mathcal{G}$, sensor/scheduler $\mathcal{S}$, and controller $\mathcal{C}$. The solid line represents an ideal channel, whereas the dashed line indicates a resource-constrained channel.}
	\label{fig:block_diagram}
\end{figure}

{\it Control architecture.}
We consider the feedback control system depicted in Figure~\ref{fig:block_diagram}. A physical plant $\mathcal{G}$, whose dynamics can be represented by a linear time-invariant stochastic system, is being controlled. A sensor $\mathcal{S}$ takes periodic samples of the plant output $\y_{k}^{}$ and transmits the estimate of the plant state $\x_{k}^{}$ to the controller over a resource-constrained communication channel. To tackle the resource constraint, the sensor employs an event-based scheduler, that makes a transmission decision by comparing its estimate of the plant state with the estimate at the controller. The controller $\mathcal{C}$ computes new control actions based on the available information. Whenever the controller receives a new state estimate from the sensor, it calculates a control command based on this state estimate. Otherwise, it runs an estimator to predict the plant state, and it uses this information to calculate a new control action. In this context, we are interested in deriving analytical performance guarantees, both regarding the control performance and the number of transmissions between the sensor and the controller.

{\it Process model.}
The system $\mathcal{G}$ is modeled as a discrete-time, linear time-invariant (LTI) system,
\begin{align}
	\x_{k+1} =\; A\x_{k} + B\bu_{k} + \w_{k}  \;, \label{eqn:stochastic_system}
\end{align}
driven by the control input $\bu_{k}\in\mathbb{R}^{m}$ (calculated by the controller $\mathcal{C}$), and an unknown  noise process $\w_{k}\in\mathbb{R}^{n}$. The state $\x_{k}\in\mathbb{R}^{n}$ is available only indirectly through the noisy output measurement
\begin{align}
		\y_{k} =\; C \x_{k} + \v_{k} \;. \label{eqn:system_output}
\end{align}
The two noise sources $\w_{k}^{}\in\mathbb{R}^{n}$ and $\v_{k}^{}\in\mathbb{R}^{p}$ are assumed to be uncorrelated zero-mean Gaussian white-noise random processes with co-variance matrices $W\in\mathbb{S}_{\succeq 0}^{n}$ and $V\in\mathbb{S}_{\succ 0}^{p}$, respectively. We refer to $\{\w_{k}^{}\}_{k\geq 0}^{}$ as the process noise, and to $\{\v_{k}^{}\}_{k\geq 0}^{}$ as the measurement noise. The initial state $\x_{0}^{}$ is modeled as a Gaussian distributed random variable with mean $\bar{\x}_{0}^{}$ and covariance $X_{0}^{}\in\mathbb{S}_{\succeq 0}^{n}$. We assume that the pairs $(A,B)$ and $(A,V^{1/2})$ are controllable while the pair $(A,C)$ is observable.

{\it Sensor, pre-processor, and scheduler.}
Using a standard Kalman filter, the sensor locally computes minimum mean squared error (MMSE) estimates $\hat{\x}_{k \mid k}^{s}$ of the plant state $\x_{k}^{}$ based on the information available to the sensor at time $k$, and transmits them to the controller. As noted in~\cite{XuH:05}, sending local state estimates, in general, provides better performance than transmitting measurements. The sensor also employs a transmission scheduler, which decides whether or not to send the current state estimate to the controller at each time-step $k\in\mathbb{N}_{0}^{}$ as determined~by
\begin{align}
	\sigma_{k}^{} =
	\begin{cases}
		1 & \text{if MMSE estimate $\hat{\x}_{k \mid k}^{s}$ is sent} \;, \\
		0 & \text{otherwise} \;.
	\end{cases}
	\label{eqn:scheduling_mechanism}
\end{align}
\begin{assumption}
  The sensor $\mathcal{S}$ has precise knowledge of the control policy used to generate control actions, which are computed by the controller and applied by the actuator to the plant. Hence, the information set of the smart sensor $\mathcal{S}$ contains all controls used up to time $k-1$.
\end{assumption}

The information set available to the sensor at time $k\in\mathbb{N}_{0}^{}$ is:
\begin{align}
	\mathcal{I}_{k}^{s} = \{ \sigma_{0}^{}, \cdots, \sigma_{k-1}^{}; \y_{0}^{}, \cdots, \y_{k}^{}; \bu_{0}^{}, \cdots, \bu_{k-1}^{} \} \;.
\end{align}
The minimum mean squared error estimate $\hat{\x}_{k \mid k}^{s}$ of the plant state $\x_{k}^{}$ can be computed recursively starting from the initial condition $\hat{\x}_{0\mid -1}^{s} = \bar{\x}_{0}^{}$ and $P_{0\mid -1}^{s} = X_{0}^{}$ using a Kalman filter \cite{MolinHirche_noisy}. At this point, it is worth reviewing the fundamental equations underlying the Kalman filter algorithm. The algorithm consists of two steps:
\begin{itemize}
  \item \textbf{Prediction step:} This step predicts the state, estimation error, and estimation error covariance at time $k$ dependent on information at time $k-1$:
  \begin{align}
  	\hat{\x}_{k \mid k-1}^{s} \triangleq &\; \mathbf{E}\big[ \x_{k}^{} \mid \mathcal{I}_{k-1}^{s} \big] 
    =\; A\hat{\x}_{k-1 \mid k-1}^{s} + B\bu_{k-1}^{} \\
  	\tilde{\x}_{k \mid k-1}^{s} \triangleq &\; \x_{k}^{} - \hat{\x}_{k \mid k-1}^{s} = A\tilde{\x}_{k-1 \mid k-1}^{s} + \w_{k-1}^{} \\
  	P_{k \mid k-1}^{s} \triangleq &\; \mathbf{E}\big[ \tilde{\x}_{k \mid k-1}^{s}\tilde{\x}_{k \mid k-1}^{s\top} \mid \mathcal{I}_{k-1}^{s} \big] 
  	 =\; AP_{k-1 \mid k-1}^{s}A_{}^{\top} + W .
  \end{align}
  \item \textbf{Update step:} This step updates the state, estimation error, and estimation error covariance using a blend of the predicted state and the observation $\y_{k}^{}$:
  \begin{align}
  	\hat{\x}_{k \mid k}^{s} \triangleq &\; \mathbf{E}\big[ \x_{k}^{} \mid \mathcal{I}_{k}^{s} \big] 
    =\; \hat{\x}_{k \mid k-1}^{s} + K_{k}^{}\big( \y_{k}^{} - C\hat{\x}_{k \mid k-1}^{s} \big) \label{eqn:correction_step_sensor} \\
  	\tilde{\x}_{k \mid k}^{s} \triangleq &\; \x_{k}^{} - \hat{\x}_{k \mid k}^{s}
    =\; (\I_{n}^{} - K_{k}^{}C)A\tilde{\x}_{k-1 \mid k-1}^{s} \nonumber\\
    & \hspace{20mm} + (\I_{n}^{} - K_{k}^{}C)\w_{k-1}^{} - K_{k}^{}\v_{k}^{} \label{eqn:correction_error_step_sensor} \\
  	P_{k \mid k}^{s} \triangleq &\; \mathbf{E}\big[ \tilde{\x}_{k \mid k}^{s}\tilde{\x}_{k \mid k}^{s\top} \mid \mathcal{I}_{k}^{s} \big] = (\I_{n}^{} - K_{k}^{}C)P_{k \mid k-1}^{s},
  \end{align}
  where the gain matrix is given by
  \begin{align}
    K_{k}^{} \triangleq &\; P_{k \mid k-1}^{s}C_{}^{\top}\big( CP_{k \mid k-1}^{s}C_{}^{\top} + V \big)_{}^{-1} \;.
  \end{align}
\end{itemize}

It is worth noting that the estimation error at the sensor $\tilde{\x}_{k \mid k}^{s}$ is Gaussian with zero-mean and co-variance $P_{k \mid k}^{s}$, that evolves according to the standard Riccati recursion~\cite[Chapter~9]{KSH:00}. Since the pair $(A,C)$ is observable and the pair $(A,W_{}^{1/2})$ is controllable, the matrices $P_{k \mid k-1}^{s}$ and $K_{k}^{}$ converge exponentially to steady state values $P_{\infty}^{s}$ and $K_{\infty}^{}$, respectively. Similarly, the matrix $P_{k \mid k}^{s}$ also converges to a steady state value, i.e., $F_{\infty}^{s} \triangleq \; (\I_{n}^{} - K_{\infty}^{}C)P_{\infty}^{s}$.

The scheduler and the sensor are collocated, and the scheduler has access to all available information at the sensor. Moreover, the scheduler employs an event-based triggering mechanism to decide if there is a need for transmission of an updated state estimate from the sensor to the controller. The occurrence of information transmission is defined as
\begin{align}
	\sigma_{k}^{} =
	\begin{cases}
		1 & \text{if $\delta_{k}^{}=1$ or $\tau_{k-1}^{} = \mathrm{T}$} \;, \\
		0 & \text{otherwise} \;,
	\end{cases}
	\label{eqn:scheduling_mechanism_2}
\end{align}
where $\delta_{k}^{}$ is a (random) binary decision variable (which in this paper evolves according to~\eqref{eqn:probabilistic_threshold_based_rule}), $\tau_{k}^{}$ is a non-negative integer variable introduced to describe the time elapsed since the last transmission, and $\mathrm{T}$ is a time-out interval. Such a time-out mechanism is critical in event-triggered control systems to guard against faulty components; see, e.g.,~\cite{DGQ+:17b}.

To maintain the Gaussianity  of the comparison error
$$\e_{k \mid k-1}^{} \triangleq \hat{\x}_{k\mid k}^{s} - \hat{\x}_{k \mid k-1}^{c},$$
(note that $\hat{\x}_{k\mid k}^{s}$ is defined in~\eqref{eqn:correction_step_sensor} while $\hat{\x}_{k\mid k-1}^{c}$ is introduced in~\eqref{eqn:prediction_step_controller}) a variant of the stochastic triggering mechanism proposed in~\cite{HMW+:15,WRH+:16,WMS+:16} is used. More specifically, the scheduler will decide to transmit a new sensor packet according to the following decision rule:
\begin{align}
	\delta_{k}^{} =
	\begin{cases}
		0 & \text{with prob.}~e_{}^{-\lambda \langle \e_{k \mid k-1}^{},\e_{k \mid k-1}^{} \rangle} \;, \\
		1 & \text{with prob.}~1 - e_{}^{-\lambda \langle \e_{k \mid k-1}^{},\e_{k \mid k-1}^{} \rangle} \;,
	\end{cases}
	\label{eqn:probabilistic_threshold_based_rule}
\end{align}
where the triggering parameter $\lambda$ is a given positive scalar. As can be seen in~\eqref{eqn:probabilistic_threshold_based_rule}, the probability of transmitting a new sensor packet (i.e., $\sigma_{k}^{}=1$) converges to one as $\lambda$ goes to infinity. In other words, for large values of $\lambda$, the communication between the sensor and the controller is more likely to be triggered.

The integer-valued random process $\{ \tau_{k}^{} \}_{k\geq 0}^{}$ in~\eqref{eqn:scheduling_mechanism_2} describes how many time instances ago the last transmission of a sensor packet occurred. Whenever a sensor packet is transmitted from the sensor to the controller, $\tau_{k}^{}$ is reset to zero. Thus, the evolution of the random process $\{ \tau_{k}^{} \}_{k\geq 0}^{}$ is defined by
\begin{align}
	\tau_{k}^{} =
	\begin{cases}
		0 & \text{if}~\delta_{k}^{} = 1~\text{or}~\tau_{k-1}^{} = \mathrm{T} \;, \\
		1 + \tau_{k-1}^{} & \text{otherwise} \;,
	\end{cases} \label{eqn:timer}
\end{align}
or equivalently,
\begin{align}
	\tau_{k}^{} =
	\begin{cases}
		0 & \text{if}~\sigma_{k}^{} = 1 \;, \\
		1 + \tau_{k-1}^{} & \text{if}~\sigma_{k}^{} = 0 \;,
	\end{cases} \label{eqn:timer_2}
\end{align}
where $\tau_{-1}^{} \triangleq 0$. Notice that the number of time steps between two consecutive transmissions is bounded by the time-out interval $\mathrm{T}<\infty$. If the number of samples since the last transmission exceeds a time-out value of $\mathrm{T}$, the sensor will attempt to transmit new data to the controller even if the comparison error does not satisfy the triggering condition~\eqref{eqn:probabilistic_threshold_based_rule}. Thus, a transmission (i.e., $\sigma_k=1$) will occur when either $\delta_k = 1$ or there is a time-out.

\begin{remark} \label{rem:equivalence}
	It is worth noting that, as can be seen in~\eqref{eqn:timer_2}, the events $\{\sigma_{k}^{}=1\}$ and $\{\tau_{k}^{}=0\}$ are equivalent to each other.
\end{remark}

At time instances when $\sigma_{k}^{}=1$, the sensor transmits its local state estimate $\hat{\x}_{k\mid k}^{s}$ to the controller. As a result, the information set available to the controller at time $k\in\mathbb{N}_{0}^{}$ (and before deciding upon $\bu_{k}^{}$) can be defined as:
\begin{align*}
	\mathcal{I}_{k}^{c} \triangleq \big\{ \sigma_{0}^{}, \cdots, \sigma_{k}^{}; \sigma_{0}^{}\hat{\x}_{0 \mid 0}^{s}, \cdots, \sigma_{k}^{}\hat{\x}_{k \mid k}^{s}; \bu_{0}^{}, \cdots, \bu_{k-1}^{} \big\} \;.
\end{align*}

Under the event-based scheduling mechanism,~\eqref{eqn:scheduling_mechanism_2} -- \eqref{eqn:timer}, the controller runs an MMSE estimator to compute estimates of the plant state $\x_{k}^{}$ as follows:
\begin{align}
	\hat{\x}_{k \mid k}^{c} & \triangleq \mathbf{E}\big[ \x_{k}^{} \mid \mathcal{I}_{k}^{c} \big] =
	\begin{cases}
		\hat{\x}_{k \mid k}^{s} & \text{if}~\sigma_{k}^{} = 1 \;, \\
		\hat{\x}_{k \mid k-1}^{c} & \text{otherwise} \;,
	\end{cases} \\
	\hat{\x}_{k \mid k-1}^{c} & \triangleq \mathbf{E}\big[ \x_{k}^{} \mid \mathcal{I}_{k-1}^{c} \big] =  A \hat{\x}_{k-1 \mid k-1}^{c} + B \bu_{k-1}^{} \;, \label{eqn:prediction_step_controller}
\end{align}
where $\hat{\x}_{k \mid k-1}^{c}$ is the optimal estimate at the controller if the sensor did not transmit any information at time-step $k\in\mathbb{N}_{0}^{}$.
Note that the optimality of this estimator can be shown by using a similar argument to that provided in~\cite[Lemma~4]{WMS+:16}.

\begin{assumption}
  In addition to computing  $\hat{\x}_{k\mid k}^{s}$, the sensor operates another estimator, which mimics the one at the controller, since transmission decisions rely on both $\hat{\x}_{k\mid k}^{s}$ and $\hat{\x}_{k\mid k-1}^{c}$; see~\eqref{eqn:probabilistic_threshold_based_rule}. This can be done provided we make the following assumption.
\end{assumption}

\begin{assumption}
	Both the smart sensor $\mathcal{S}$ and the controller $\mathcal{C}$ know the plant model $\mathcal{G}$ (but not realizations of the noise processes).
\end{assumption}

{\it Controller design and performance criterion.}
We aim at finding the control strategies $\bu_{k}^{}$, as a function of the admissible information set $\mathcal{I}_{k}^{c}$, to minimize a quadratic cost function of the form
\begin{align}
	J_{N} = \mathbf{E}\bigg[ \x_{N}^{\top}Q_{f}\x_{N}^{} + \sum_{k=0}^{N-1}\Big( \x_{k}^{\top}Q\x_{k}^{} + \bu_{k}^{\top}R\bu_{k}^{} \Big) \bigg] \;,
	\label{eqn:control_loss_function}
\end{align}
where $Q, Q_{f}^{}\in\mathbb{S}_{\succeq 0}^{n}$ and $R\in\mathbb{S}_{\succ 0}^{m}$. At time instances when $\sigma_{k}^{}=1$ (i.e., the controller has received sensor packets), the controller uses the state estimate $\hat{\x}_{k\mid k}^{s}$ which is transmitted by the sensor. However, at time instances when $\sigma_{k}^{}=0$, the controller uses the outcome of the estimator at the controller side. As is well-known in related situations (see e.g.,~\cite{MoH:13}), if the transmission decision $\sigma_{k}^{}$ is independent of the control strategy $\bu_{k}^{}$, then the certainty equivalent controller is optimal. In Section~\ref{sec:main_results}, we will confirm that the certainty equivalent controller is optimal under the event-based scheduler proposed above.

\section{Main Results}\label{sec:main_results}
We wish to quantify the communication rate and control performance of the feedback control system described by~\eqref{eqn:stochastic_system} and~\eqref{eqn:system_output}, where the event-based triggering mechanism~\eqref{eqn:probabilistic_threshold_based_rule} determines the communication between the sensor and the controller. We will first demonstrate that the time elapsed between two consecutive transmissions can be regarded as a discrete-time, finite state, time-homogeneous Markov chain. Then, using an ergodicity property, we will provide an analytical formula for the communication rate between the sensor and the controller. Subsequently, we will show that the certainty equivalent controller is still optimal with the event-triggering rule~\eqref{eqn:probabilistic_threshold_based_rule}. Lastly, we will compute the control performance analytically for the infinite horizon case.

\begin{assumption}
\label{steady_state_KF_assumption}
	In the rest of this paper, we will assume that the local Kalman filter at the sensor runs in steady state.
\end{assumption}

We first define the state prediction error at the controller
\begin{align}
	\tilde{\x}_{k \mid k-1}^{c} \triangleq \x_{k}^{} - \hat{\x}_{k \mid k-1}^{c} \;,
\end{align}
which evolves as
\begin{align}
	\tilde{\x}_{k+1 \mid k}^{c} =
	\begin{cases}
		A\tilde{\x}_{k \mid k}^{s}	 +\w_{k}^{} & \text{if} ~ \sigma_{k}^{} = 1 \;, \\
		A\tilde{\x}_{k \mid k-1}^{c} + \w_{k}^{} & \text{if} ~ \sigma_{k}^{} = 0 \;.
	\end{cases}
\end{align}
Then, we  define the state estimation error at the controller
\begin{align}
	\tilde{\x}_{k \mid k}^{c} \triangleq \x_{k}^{} - \hat{\x}_{k \mid k}^{c} \;,
\end{align}
which evolves as
\begin{align}
	\tilde{\x}_{k \mid k}^{c} =
	\begin{cases}
		\tilde{\x}_{k \mid k}^{s}	& \text{if} ~ \sigma_{k}^{} = 1 \;, \\
		A\tilde{\x}_{k-1 \mid k-1}^{c} + \w_{k-1}^{} & \text{if} ~ \sigma_{k}^{} = 0 \;.
	\end{cases}
\end{align}
Define also the comparison errors:
\begin{align}
	\e_{k \mid k-1}^{} \triangleq &\; \hat{\x}_{k \mid k}^{s} - \hat{\x}_{k \mid k-1}^{c} = \tilde{\x}_{k \mid k-1}^{c} - \tilde{\x}_{k \mid k}^{s} \;, \label{eqn:error_1} \\
	\e_{k \mid k}^{} \triangleq &\; \hat{\x}_{k \mid k}^{s} - \hat{\x}_{k \mid k}^{c} = \tilde{\x}_{k \mid k}^{c} - \tilde{\x}_{k \mid k}^{s} \;. \label{eqn:error_2}
\end{align}

Whenever a transmission occurs (i.e., $\tau_{k}^{}=0$), the state estimation error $\tilde{\x}_{k\mid k}^{c}$ at the controller is equal to $\tilde{\x}_{k\mid k}^{s}$, since the most recent sensor packet is available at the controller. It is then possible to write the stochastic recurrence equations~\eqref{eqn:error_1} and~\eqref{eqn:error_2} as
\begin{align}
	\e_{k+1 \mid k}^{} =
	\begin{cases}
		\boldeta_{k}^{} & \text{if}~\tau_{k}^{} = 0 \;, \\
		A\e_{k \mid k-1}^{} + \boldeta_{k}^{} & \text{if}~\tau_{k}^{} \neq 0 \;,
	\end{cases}
	\label{eqn:error_3}
\end{align}
and
\begin{align}
	\e_{k \mid k}^{} =
	\begin{cases}
		0 & \text{if}~\tau_{k}^{} = 0 \;, \\
		A\e_{k-1 \mid k-1}^{} + \boldeta_{k-1}^{} & \text{if}~\tau_{k}^{} \neq 0 \;,
	\end{cases}
	\label{eqn:error_4}
\end{align}
where $\boldeta_{k}^{} \triangleq K_{\infty}^{}\big(C(A\tilde{\x}_{k\mid k}^{s} + \w_{k}^{}) + \v_{k+1}^{}\big)$. Notice that the comparison errors $\e_{k \mid k-1}^{}$ and $\e_{k \mid k}^{}$ propagate according to a linear system with open-loop dynamics $A$, driven by the  process $\boldeta_{k}^{}$.

\begin{lemma}\label{lem:white_noise}
	$\{ \boldeta_{k}^{} \}_{k\geq 0}^{}$ is a sequence of pairwise independent Gaussian random vectors such that $\boldeta_{k}^{}\sim\mathcal{N}(0,\Pi_{\eta}^{})$ with $\Pi_{\eta}^{} \triangleq K_{\infty}^{}CP_{\infty}^{s}$.
\end{lemma}

\begin{remark}
	If the sensor has perfect state measurements (i.e., $\y_{k}^{} = \x_{k}^{}$), then $\boldeta_{k}^{}$ is equal to $\w_{k}^{}$.
\end{remark}

\begin{definition}[Cumulative error] \label{def:cumulative_error}
We shall characterize the cumulative comparison error (i.e., the error that occurs in estimation at the controller over time due to intermittent transmissions) via
\begin{align}
	\beps_{k}^{}(i) \triangleq \sum_{j=0}^{i} A_{}^{j} \boldeta_{k-j}^{} \;. \label{eqn:random_variable}
\end{align}
\end{definition}

Using Definition~\ref{def:cumulative_error}, the stochastic recurrence equations~\eqref{eqn:error_3} and \eqref{eqn:error_4} can be then re-written as
\begin{align}
	\e_{k+1 \mid k}^{} =&\; \beps_{k}^{}(\tau_{k}^{} ) \;, \label{eqn:error_5} \\
	\e_{k \mid k}^{} =&\;
  \begin{cases}
    0 & \text{if} ~ \tau_{k}^{} = 0 \;, \\
    \beps_{k-1}^{}(\tau_{k-1}^{}) & \text{if} ~ \tau_{k}^{} \neq 0 \;.
  \end{cases}  \label{eqn:error_6}
\end{align}
%

\begin{lemma}[Markov process] \label{lem:markov_chain}
The random process $\{ \tau_{k}^{} \}_{k\geq 0}^{}$ is an ergodic, time-homogeneous Markov chain with a finite state space $\mathcal{B} \triangleq \{ 0,1, \cdots, \mathrm{T} \}$. Thus, it has a unique invariant distribution $\boldsymbol{\pi} \triangleq \big[ \pi(0) ~ \pi(1) ~ \cdots ~ \pi(\mathrm{T}) \big]\in\mathbb{R}_{}^{1\times\mathrm{T}+1}$ such that $\sum_{i\in\mathcal{B}}^{}\pi(i) = 1$ and $\pi(i)>0$ for all $i\in\mathcal{B}$.
\end{lemma}

\begin{lemma}[Augmented cumulative error vector]
Consider $\bar{\beps}_{k}^{}(i) \triangleq \big[ \beps_{k}^{\top}(0) ~ \beps_{k+1}^{\top}(1) ~ \cdots ~ \beps_{k+i}^{\top}(i) \big]_{}^{\top}$ with $\beps_{k}^{}(i)$ as in~\eqref{eqn:random_variable}. Then, $\bar{\beps}_{k}^{}(i)$ is a random vector having a multivariate normal distribution with zero-mean and co-variance:
\begin{align*}
	\Sigma_{\epsilon}^{}(i) \triangleq
	\begin{bmatrix}
		\Pi_{\eta}^{} 	& \Pi_{\eta}^{}A_{}^{\top} 															& \hdots & \Pi_{\eta}^{}(A_{}^{i})_{}^{\top} \\
		\star 				& \sum\limits_{j=0}^{1}A_{}^{j}\Pi_{\eta}^{}(A_{}^{j})_{}^{\top}  	& \hdots & \sum\limits_{j=0}^{1}A_{}^{j}\Pi_{\eta}^{}(A_{}^{j+i-1})_{}^{\top} \\
		\vdots				& \vdots 																									& \ddots & \vdots \\
		\star 				& \star 																									& \hdots & \sum\limits_{j=0}^{i}A_{}^{j}\Pi_{\eta}^{}(A_{}^{j})_{}^{\top}
\end{bmatrix}
\end{align*}
for any $i\in\{ 0, 1, \cdots, \mathrm{T}-1 \}$.
\end{lemma}

The next lemma computes the transition probabilities of the Markov chain defined in Lemma~\ref{lem:markov_chain}.

\begin{lemma}[Transition probabilities]\label{lem:transition_probabilities}
The transition matrix of the Markov chain $\{ \tau_{k}^{} \}_{k\geq 0}^{}$ is given by
\begin{align*}
	P_{\lambda}^{} =
	\begin{bmatrix}
		p_{00}^{} & 1 - p_{00}^{} & 0                   & \hdots & 0 \\
		p_{10}^{} & 0                   & 1 - p_{10}^{} & \hdots & 0\\
		\vdots       & \vdots           & \vdots & \ddots & \vdots  \\
		p_{\mathrm{T}-1,0}^{} & 0                   & 0                   & \hdots & 1 - p_{\mathrm{T}-1,0}^{} \\
		p_{\mathrm{T},0}^{}				 & 0                   & 0 & \hdots & 0
	\end{bmatrix} ,
\end{align*}
where the non-zero transition probabilities are computed as
\begin{equation*}
p_{ij} =
\begin{cases}
1 - \dfrac{1}{\sqrt{\vert \I_{n}^{} + 2\lambda\Sigma_{\epsilon}^{}(0) \vert}}  & \text{if}~i = 0, ~ j=0, \\
1 - \sqrt{\dfrac{\vert \I_{in}^{} + 2\lambda\Sigma_{\epsilon}^{}(i-1) \vert}{\vert \I_{(i+1)n}^{} + 2\lambda\Sigma_{\epsilon}^{}(i) \vert}}  & \hspace{-2mm}\begin{array}{l} \text{if}~i\in\{1,\cdots,\mathrm{T}-1\}, \\ ~~ j=0, \end{array} \\
1 - p_{i0} & \hspace{-2mm}\begin{array}{l} \text{if}~i\in\{0,\cdots,\mathrm{T}-1\}, \\ ~~ j=i+1, \end{array} \\
1 & \text{if}~i = \mathrm{T}, ~ j = 0 \;.
\end{cases}
\end{equation*}
\end{lemma}

The visit of the Markov chain $\{ \tau_{k}^{} \}_{k\geq 0}^{}$ to the state $0$ is analogous to a transmission (i.e., $\sigma_{k}^{} = 1$) of the estimate of the plant state $\x_{k}^{}$ from the sensor to the controller. Using Remark~\ref{rem:equivalence} and the ergodic theorem for Markov chains~\cite[Theorem~5.3]{Bre:99}, we have:
\begin{align}
	\bar{\sigma} = \lim\limits_{n\rightarrow\infty}\frac{1}{n}\sum_{k=0}^{n-1}\sigma_{k}^{} \Longleftrightarrow \pi(0) = \lim\limits_{n\rightarrow\infty}\frac{1}{n}\sum_{k=0}^{n-1}\mathbbm{1}_{\{\tau_{k}^{}=0\}} \;,
\end{align}
where $\pi(0)$ is the empirical frequency of transmissions. With the transition probabilities of this Markov chain, we can give an explicit characterization of  the average communication rate of the event-triggered control system:

\begin{theorem}[Communication rate]\label{thm:communication_rate}
	The average communication rate between the sensor and the controller under the stochastic event-based triggering mechanism, proposed in~\eqref{eqn:scheduling_mechanism_2} -- \eqref{eqn:timer}, for a fixed $\lambda > 0$ is given by
	\begin{align}
		\bar{\sigma} = \frac{1}{1 + \sum_{n=1}^{\mathrm{T}}\prod_{m=0}^{n-1}(1 - p_{m0}^{})} \;. \label{eqn:communication_rate}
	\end{align}
\end{theorem}

\begin{remark}
	Note that, as $\lambda$ goes to zero, the communication between the sensor and the controller becomes periodic.
\end{remark}

The next theorem describes the optimal control law for the event-triggered control system at hand.

\begin{theorem}[Optimal control law]\label{thm:optimal_control_law}
	Consider the system \eqref{eqn:stochastic_system} and \eqref{eqn:system_output}, and the problem of minimizing the cost function~\eqref{eqn:control_loss_function} under the event-based triggering mechanism~\eqref{eqn:scheduling_mechanism_2} -- \eqref{eqn:timer} for a fixed $\lambda > 0$. Then, there exists a unique admissible optimal control policy
	\begin{align}
		\bu_{k}^{} = - L_{k}^{}\mathbf{E}\big[ \x_{k}^{} \mid \mathcal{I}_{k}^{c} \big] = - L_{k}^{}\hat{\x}_{k \mid k}^{c} \;,
	\end{align}
	where
	\begin{align}
		L_{k}^{} =&\; (B_{}^{\top}S_{k+1}^{}B + R)_{}^{-1} B_{}^{\top}S_{k+1}^{}A , \label{eqn:optimal_control_gain} \\
		S_{k}^{} =&\; A_{}^{\top} S_{k+1}^{} A + Q \nonumber\\
		& - A_{}^{\top}S_{k+1}^{}B (B_{}^{\top}S_{k+1}^{}B + R)_{}^{-1} B_{}^{\top}S_{k+1}^{}A , \label{eqn:riccati_equation}
	\end{align}
	with initial values $S_{N}^{} = Q_{f}^{}$. The minimum value of the cost function is obtained as
	\begin{multline}
		J_{N}^{} = \bar{\x}_{0}^{\top}S_{0}^{}\bar{\x}_{0}^{} + \textnormal{Tr}\big( S_{0}^{}X_{0}^{} \big) + \sum_{k=0}^{N-1}\textnormal{Tr}\big( S_{k+1}^{}W \big) \\
		+ \sum_{k=0}^{N-1} \textnormal{Tr}\big( P_{k \mid k}^{s}M_{k}^{} \big) + \sum_{k=0}^{N-1} \mathbf{E}\Big[ \e_{k \mid k}^{\top} M_{k}^{} \e_{k \mid k}^{} \Big] \;. \label{eqn:inf_expected_cost}
	\end{multline}
	where $M_{k}^{} \triangleq L_{k}^{\top}(B_{}^{\top}S_{k+1}^{}B + R)L_{k}^{}$.
\end{theorem}

\begin{remark}
  Our result should be viewed in the light of the limited information available to the controller. At every time step $k\in\mathbb{N}_{0}^{}$, the controller computes an optimal control input based on the information set $\mathcal{I}_{k}^{c}$. Our result is akin to the one derived in~\cite{MoH:13}, however here we can also provide the closed-form expression of the control cost for the infinite horizon case (see Theorem~\ref{thm:control_performance}).
\end{remark}

\begin{lemma}[Gaussianity-preservation] \label{lem:Gaussianity_preservation}
The conditional random variable, $\e_{k \mid k}^{} \mid \tau_{k}^{}=i$, has a Gaussian distribution with zero-mean and co-variance:
\begin{align*}
	\Sigma_{e}^{}(0) &= \0_{n}^{} \;, \\
	\Sigma_{e}^{}(i) &= \frac{1}{2\lambda}\I_{n}^{} - \frac{1}{4\lambda_{}^{2}}\bigg( A\Sigma_{e}^{}(i-1)A_{}^{\top} + \Pi_{\eta}^{} + \frac{1}{2\lambda}\I_{n}^{} \bigg)_{}^{-1}.
\end{align*}
%
%
\end{lemma}

Using the previous theorems, we have the following result to calculate the average control performance measured by a linear-quadratic function.

\begin{theorem}[Infinite horizon control performance] \label{thm:control_performance}
 	Suppose the pairs $(A,B)$ and $(A,W_{}^{1/2})$ are controllable, and the pairs $(A,C)$ and $(A,Q_{}^{1/2})$ are observable. Moreover, suppose that $\lambda > 0$. Then, we have the following:
 	\begin{itemize}
 	\item[(a)] The infinite horizon optimal controller gain is constant:
 	\begin{align}
 		L_{\infty}^{} \triangleq \lim_{k\rightarrow\infty}^{} L_{k}^{} = (B_{}^{\top}S_{\infty}^{}B + R)_{}^{-1} B_{}^{\top}S_{\infty}^{}A \;. \label{eqn:stationary_optimal_control_gain}
 	\end{align}
 	\item[(b)] The matrices $S_{\infty}^{}$ and $P_{\infty}^{s}$ are the positive definite solutions of the following algebraic Riccati equations:
	\begin{align}
		S_{\infty}^{} \triangleq &\; A_{}^{\top} S_{\infty}^{} A + Q \nonumber\\
		&  - A_{}^{\top}S_{\infty}^{}B (B_{}^{\top}S_{\infty}^{}B + R)_{}^{-1} B_{}^{\top}S_{\infty}^{}A \;, \label{eqn:algebraic_riccati_equation_1} \\
		P_{\infty}^{s} \triangleq &\; A P_{\infty}^{s} A_{}^{\top} + W \nonumber\\
		&  - AP_{\infty}^{s}C_{}^{\top} (C P_{\infty}^{s}C_{}^{\top} + V)_{}^{-1} CP_{\infty}^{s}A_{}^{\top} \;. \label{eqn:algebraic_riccati_equation_2}
	\end{align}
 	\item[(c)] The expected minimum cost converges to the following value:
	\begin{align}
		J_{\infty}^{} \triangleq &\lim_{N\rightarrow\infty}^{}\frac{1}{N}J_{N}^{} = \textnormal{Tr}(S_{\infty}^{}W) + \textnormal{Tr}(F_{\infty}^{s}M_{\infty}^{}) \nonumber\\
		& \hspace{21mm} + \sum_{i=1}^{\mathrm{T}}\pi(i) \textnormal{Tr}\Big( M_{\infty}^{} \Sigma_{e}^{}(i) \Big), \label{eqn:contrl_performance_infinite_horizon}
	\end{align}	
	where $F_{\infty}^{s} \triangleq (\I_{n}^{} - K_{\infty}^{}C)P_{\infty}^{s}$, $M_{\infty}^{} \triangleq L_{\infty}^{\top}(B_{}^{\top}S_{\infty}^{}B + R)L_{\infty}^{}$, and $\boldsymbol{\pi} = \big[ \pi(0) ~ \pi(1) ~ \cdots ~ \pi(\mathrm{T}) \big]$ satisfies $\boldsymbol{\pi} = \boldsymbol{\pi}P_{\lambda}^{}$.
 	\end{itemize}
\end{theorem}

\section{Numerical Example}\label{sec:numerical_example}
In this section, numerical simulations are provided to assess the performance of the stochastic event-triggering algorithm proposed in Section~\ref{sec:problem_formulation}, and verify the theoretical results presented in Section~\ref{sec:main_results}. To this end, the system parameters are chosen as follows:
\begin{align*}
	A =& \begin{bmatrix} 1.2 & 1 \\ 0 & 0.9\end{bmatrix}, ~
	B = \begin{bmatrix} 0 \\ 1 \end{bmatrix}, ~
	C = \begin{bmatrix} 1 & 0 \end{bmatrix}, ~ V = 1, \\
	X_{0}^{} =&\; W = \begin{bmatrix} 1 & 0.5 \\ 0.5 & 1 \end{bmatrix}, ~
	Q = \begin{bmatrix} 2 & 0.5 \\ 0.5 & 2 \end{bmatrix}, ~
	R = 1.
\end{align*}
The matrix $A$ has one stable (i.e., $0.9$) and one unstable eigenvalue (i.e., $1.2$). The time-out interval is set to $\mathrm{T} = 50$. Notice that the pairs $(A,B)$ and $(A,Q_{}^{1/2})$ are controllable, the pairs $(A,C)$ and $(A,W_{}^{1/2})$ are observable, and $R>0$, as required by the assumptions of the theorems presented in Section~\ref{sec:main_results}.

\begin{figure}[!t]\centering
	\includegraphics[scale=1.0]{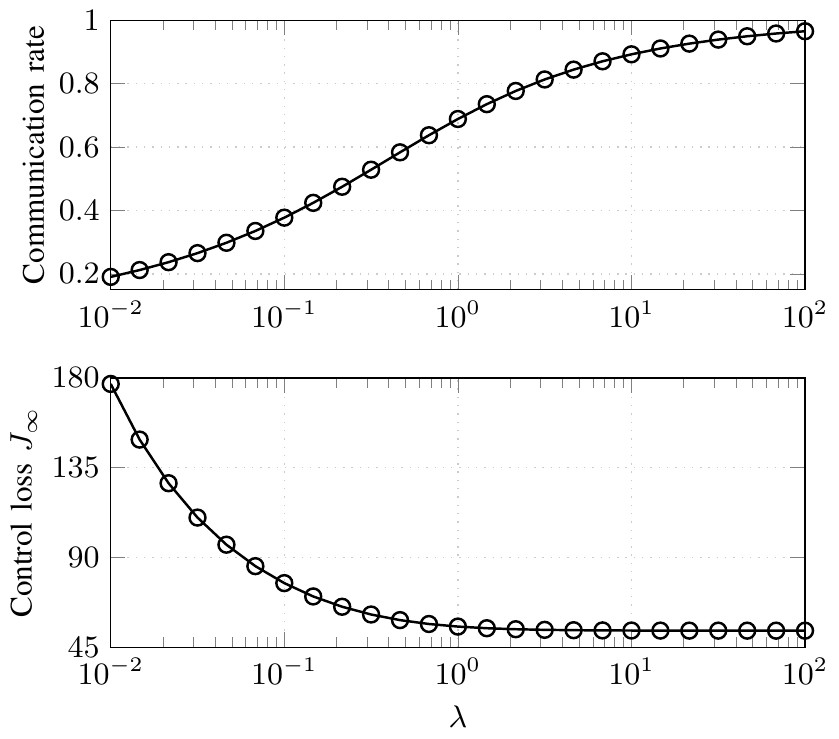}
  \caption{A comparison of transmission rate (resp. control performance) derived from the analytic expression~\eqref{eqn:communication_rate} (resp.~\eqref{eqn:contrl_performance_infinite_horizon}) and Monte Carlo simulations.} 		\label{fig:numerical_simulation_1}
\end{figure}

For various values of $\lambda$ ranging from $0.01$ to $100$, we evaluate the communication rate and the control performance as predicted by Theorems~\ref{thm:communication_rate} and~\ref{thm:control_performance}, respectively. We compare the analytic results to Monte Carlo simulations of the closed-loop system. For each value of $\lambda$, we conduct $25,000$ Monte Carlo simulations for the horizon length of $10,000$ samples, and obtain the mean communication rate and the control performance. The comparison is shown in Fig.~\ref{fig:numerical_simulation_1} for the communication rate and the control performance. It can be seen that the analytic results match the Monte Carlo simulations very closely.

\begin{figure}[!t]\centering
	\includegraphics[scale=1.0]{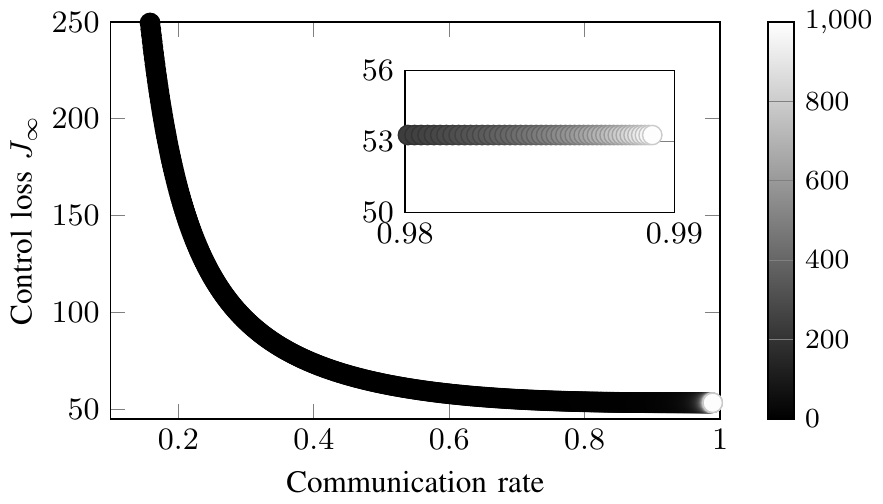}
  	\caption{The trade-off between the communication rate and the control performance (the scheduling parameter $\lambda$ is illustrated by gray scale).}
  	\label{fig:numerical_simulation_2}
\end{figure}

We can also obtain results on when changing the scheduling parameter $\lambda$ has the most effect as demonstrated in Fig.~\ref{fig:numerical_simulation_2}. There are two extreme cases: 1) as $\lambda$ goes to infinity, the communication rate becomes one, and the control performance converges to $\textnormal{Tr}(S_{\infty}^{}W) + \textnormal{Tr}(F_{\infty}^{s}M_{\infty}^{}) \approx 53.23$; 2) as $\lambda$ goes to zero, the transmission rate converges to zero, and the control performance becomes unbounded. We observe, for instance, that changing the scheduling parameter $\lambda$ from one to infinity has minimal effect on the control performance, but nearly doubles the communication frequency.  As can be seen in Fig.~\ref{fig:numerical_simulation_1}, by setting $\lambda = 1$, we can reduce the communication between the sensor and the controller by almost $40\%$, while only slightly sacrificing the control performance of the closed-loop system.

%
%
%

\section{Conclusions and Discussions}\label{sec:conclusions}
This paper has focused on the optimal control of a linear stochastic system, where a stochastic event-based scheduling mechanism governs the communication between the sensor and the controller. The scheduler is co-located at the sensor and employs a local Kalman filter. Based on the prediction error, the scheduler decides whether or not to send a new state estimate to the controller. The use of this transmission strategy reduces the communication burden in the channel. We showed that, in this setup, the optimal controller is the certainty-equivalent controller since the measurement quality is not affected by the control policy.  We also provided analytical expressions to quantify the trade-off between the communication rate and the control performance.

\begin{figure}[!t]\centering
	\includegraphics[scale=1.0]{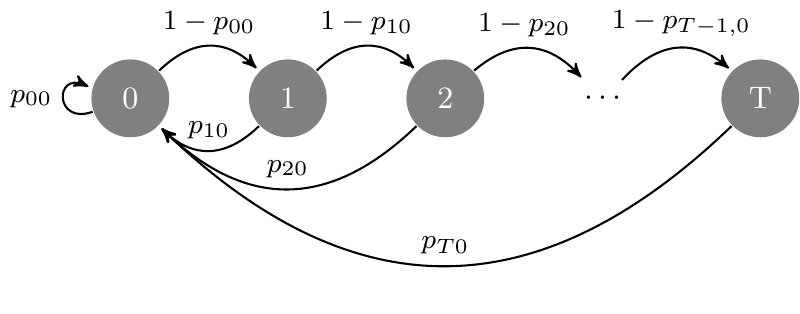}
    \caption{Transition graph of the Markov chain $\{ \tau_{k}^{} \}_{k\geq 0}^{}$.} \label{fig:Markov_Chain}
\end{figure}
%

\section{Appendix: Proofs}

{\it Proof of Lemma~\ref{lem:white_noise}:}
By Assumption~\ref{steady_state_KF_assumption}, the Kalman filter has reached its steady state. Consequently, the Kalman gain $K_{k}^{}$ and the error co-variance matrices, $P_{k \mid k-1}^{s}$ and $P_{k \mid k}^{s}$, become constant, i.e., $K_{\infty}^{}$, $P_{\infty}^{s}$, and $F_{\infty}^{s}=(\I_{n}^{} - K_{\infty}^{}C)P_{\infty}^{s}$, respectively. Let us define the following random process:
\begin{align*}
	\boldeta_{k}^{} = K_{\infty}^{}CA\tilde{\x}_{k \mid k}^{s} + K_{\infty}^{}C\w_{k}^{} + K_{\infty}^{}\v_{k+1}^{}\;.
\end{align*}
Since $\tilde{\x}_{k\mid k}^{s}$, $\w_{k}^{}$ and $\v_{k+1}^{}$ are mutually independent Gaussian vectors with zero-mean and co-variances $F_{\infty}^{s}$, $W$, and $V$, respectively, $\boldeta_{k}^{}$ is Gaussian with zero-mean and co-variance:
\begin{align}
	\Pi_{\boldeta}^{} =& \mathbf{E}\big[ \boldeta_{k}^{}\boldeta_{k}^{\top} \big] = \mathbf{E}\Big[ \big(K_{\infty}^{}CA\tilde{\x}_{k \mid k}^{s} + K_{\infty}^{}C\w_{k}^{} + K_{\infty}^{}\v_{k+1}^{}\big) \nonumber\\
	&\times \big(K_{\infty}^{}CA\tilde{\x}_{k\mid k}^{s} + K_{\infty}^{}C\w_{k}^{} + K_{\infty}^{}\v_{k+1}^{}\big)_{}^{\top} \Big] \nonumber \\
	=&\; K_{\infty}^{}\big( C(AF_{\infty}^{s}A_{}^{\top} + W)C_{}^{\top} + V \big)K_{\infty}^{\top} \nonumber \\
	\stackrel{(a)}{=}&\;  K_{\infty}^{}\big( CP_{\infty}^{s}C_{}^{\top} + V \big)K_{\infty}^{\top} \stackrel{(b)}{=} K_{\infty}^{}CP_{\infty}^{s} \;, \label{eqn:sigma_eta_relation}
\end{align}
where ($a$) is derived by writing $P_{\infty}^{s} \triangleq AF_{\infty}^{s}A_{}^{\top} + W$ while ($b$) is obtained by replacing $K_{\infty}^{\top}$ with $\big( CP_{\infty}^{s}C_{}^{\top} + V \big)_{}^{-1}CP_{\infty}^{s}$.

Since $\{ \boldeta_{k}^{} \}_{k\geq 0}^{}$ are Gaussian random vectors, pairwise independence is equivalent to
\begin{align*}
	\mathbf{E}\big[ \boldeta_{k}^{}\boldeta_{l}^{\top} \big] = \underline{\0}_{n}^{}, \qquad 0\leq k < l < \infty \;.
\end{align*}
For $k < l$, we have:
\begin{align*}
	&\mathbf{E}\big[ \boldeta_{k}^{}\boldeta_{l}^{\top} \big] =\; \mathbf{E}\Big[ \big(K_{\infty}^{}CA\tilde{\x}_{k \mid k}^{s} + K_{\infty}^{}C\w_{k}^{} + K_{\infty}^{}\v_{k+1}^{}\big) \\
	&\times\big(K_{\infty}^{}CA\tilde{\x}_{l \mid l}^{s} + K_{\infty}^{}C\w_{l}^{} + K_{\infty}^{}\v_{l+1}^{}\big)_{}^{\top} \Big] \\
	\stackrel{(a)}{=}&\; \mathbf{E}\Big[ \big(K_{\infty}^{}CA\tilde{\x}_{k \mid k}^{s} + K_{\infty}^{}C\w_{k}^{} + K_{\infty}^{}\v_{k+1}^{}\big)\tilde{\x}_{l \mid l}^{s\top} \Big] A_{}^{\top}C_{}^{\top}K_{\infty}^{\top} \\
	\stackrel{(b)}{=}&\; \Big(K_{\infty}^{}C\big( AF_{\infty}^{s}A_{}^{\top} + W \big)(\I_{n}^{} - K_{\infty}^{}C)_{}^{\top} \\
	& - K_{\infty}^{}VK_{\infty}^{\top}\Big)\Big((A-K_{\infty}^{}CA)_{}^{l-k-1}\Big)_{}^{\top}A_{}^{\top}C_{}^{\top}K_{\infty}^{\top} \\
	\stackrel{(c)}{=}&\; \Big(K_{\infty}^{}CP_{\infty}^{s}(\I_{n}^{} - K_{\infty}^{}C)_{}^{\top} - K_{\infty}^{}VK_{\infty}^{\top}\Big) \\
	& \times\Big((A-K_{\infty}^{}CA)_{}^{l-k-1}\Big)_{}^{\top}A_{}^{\top}C_{}^{\top}K_{\infty}^{\top} \\
	=&\; \Big(K_{\infty}^{}CP_{\infty}^{s} - K_{\infty}^{}(CP_{\infty}^{s}C_{}^{\top} + V)K_{\infty}^{\top}\Big) \\
	&\times\Big((A-K_{\infty}^{}CA)_{}^{l-k-1}\Big)_{}^{\top}A_{}^{\top}C_{}^{\top}K_{\infty}^{\top} \stackrel{(d)}{=} \underline{\0}_{n}^{} \;,
\end{align*}
where ($a$) holds since $\w_{l}^{}$ and $\v_{l+1}^{}$ are independent of $\hat{\x}_{k\mid k}^{s}$, $\w_{k}^{}$ and $\v_{k+1}^{}$; ($b$) is obtained by replacing $\hat{\x}_{l\mid l}^{s}$ with~\eqref{eqn:correction_error_step_sensor} iteratively from $l$ to $k$ and using the fact that $\w_{k}^{}$ and $\v_{k+1}^{}$ are independent of $\{\w_{k+1}^{},\cdots,\w_{l-1}^{}\}$ and $\{\v_{k+2}^{},\cdots,\v_{l}^{}\}$; ($c$) is obtained by writing $P_{\infty}^{s} \triangleq AF_{\infty}^{s}A_{}^{\top} + W$; and ($d$) follows from~\eqref{eqn:sigma_eta_relation}. This concludes the proof.
$\hfill\blacksquare$

{\it Proof of Lemma~\ref{lem:markov_chain}:}
For simplicity, we will use a slight abuse of notation and write $\e_{k+1}^{} = \e_{k+1\mid k}^{}$. We begin by proving that the process $\{ \tau_{k} ^{} \}_{k\geq 0}^{}$ is a Markov chain. Using the total law of probabilities and the fact that $\e_{k+1}^{}\in\mathbb{R}_{}^{n}$, we have:
\begin{align*}
	\mathbf{P}\big( \tau_{k+1}^{} \mid &~\tau_{k}^{}, \tau_{k-1}^{}, \cdots, \tau_{0}^{} \big) \\
	&=\; \int_{\mathbb{R}_{}^{n}}\mathbf{P}\big( \tau_{k+1}^{}, \e_{k+1}^{} \mid \tau_{k}^{}, \tau_{k-1}^{}, \cdots, \tau_{0}^{} \big) d\e_{k+1}^{} \\
	&\stackrel{(a)}{=}\; \int_{\mathbb{R}_{}^{n}}\mathbf{P}\big( \tau_{k+1}^{} \mid \e_{k+1}^{}, \tau_{k}^{}, \tau_{k-1}^{}, \cdots, \tau_{0}^{} \big) \\
	& \hspace{16mm}\times\mathbf{P}\big( \e_{k+1}^{} \mid \tau_{k}^{}, \tau_{k-1}^{}, \cdots, \tau_{0}^{} \big) d\e_{k+1}^{} \\
	&\stackrel{(b)}{=}\; \int_{\mathbb{R}_{}^{n}}\mathbf{P}\big( \tau_{k+1}^{} \mid \e_{k+1}^{}, \tau_{k}^{} \big)\mathbf{P}\big( \e_{k+1}^{} \mid \tau_{k}^{} \big) d\e_{k+1} \\
	&\stackrel{(c)}{=}\; \int_{\mathbb{R}_{}^{n}}\mathbf{P}\big( \tau_{k+1}^{}, \e_{k+1}^{} \mid \tau_{k}^{} \big) d\e_{k+1}^{} \\
	&=\; \mathbf{P}\big( \tau_{k+1} \mid \tau_{k}\ \big) \;,
\end{align*}
where ($a$) and ($c$) come from the definition of  conditional probability, and ($b$) holds since $\e_{k+1}$ depends stochastically only on $\tau_{k}^{}$ as described in~\eqref{eqn:error_5}, and $\tau_{k+1}^{}$ depends on $\e_{k+1}^{}$ and $\tau_{k}^{}$ as described in~\eqref{eqn:timer}. Bear in mind that knowing $\tau_{k}^{}=j$ implies knowing $\tau_{k}^{}=j, \tau_{k-1}^{}=j-1,\cdots, \tau_{k-j}^{}=0$. Consequently, the process $\{ \tau_{k}^{} \}_{k\geq 0}^{}$ is a Markov chain.

We now show the ergodicity of this Markov chain. Since the Markov chain $\{ \tau_{k}^{} \}_{k\geq 0}^{}$, depicted in Fig.~\ref{fig:Markov_Chain}, has positive transition probabilities for any $\lambda>0$, the chain is evidently irreducible. The chain is also aperiodic because the state $0$ has a non-zero probability of being reached for any $\lambda>0$. By~\cite[Theorem~3.3]{Bre:99}, this irreducible chain with finite state space $\mathcal{B}$ is positive recurrent. Since the process $\{ \tau_{k}^{} \}_{k\geq 0}^{}$ is irreducible, aperiodic and positive recurrent, it is also ergodic. As the process $\{ \tau_{k}^{} \}_{k\geq 0}^{}$ is an irreducible aperodic Markov chain with finitely many states, it has a unique invariant distribution $\boldsymbol{\pi}$ such that $\boldsymbol{\pi} P_{\lambda}^{}=\boldsymbol{\pi}$ and $\boldsymbol{\pi}\underline{\mathbf{1}}_{\mathrm{T}+1}^{}=1$; see~\cite[Corollary~2.11]{Cin:75}. This concludes the proof.
$\hfill\blacksquare$

\begin{lemma} \label{lem:probability}
Suppose that $\zeta_{k+1}^{},\zeta_{k+2}^{}, \cdots, \zeta_{k+i}^{}$ is a sample of $\zeta \stackrel[]{\text{i.i.d.}}{\sim}\mathrm{Uni}(0,1)$. Define the following events:
\begin{multline}
	\mathcal{E}_{i}^{} \triangleq \big\{ \delta_{k+1}^{}=0, \cdots, \delta_{k+i}^{}=0 \big\} \\
	= \bigcap_{j=0}^{i-1} \Big\{ \zeta_{k+j+1}^{} \leq e_{}^{-\lambda \langle \beps_{k+j}^{}(j), \beps_{k+j}^{}(j) \rangle} \Big\}
\end{multline}
for all $i\in\{ 1, 2, \cdots, \mathrm{T} \}$, with the convention that $\mathcal{E}_{0}^{}$ is a sure event. For any given $\lambda > 0$, the probability of these events $\mathcal{E}_{i}^{}$, for all $i\in\{ 1, 2, \cdots, \mathrm{T} \}$, can be computed as:
\begin{align}
	\mathbf{P}\big( \mathcal{E}_{i}^{} \big) = \frac{1}{\sqrt{\vert \I_{in}^{} + 2\lambda\Sigma_{\epsilon}^{}(i-1) \vert}}.
\end{align}
\end{lemma}

{\it Proof of Lemma~\ref{lem:probability}:}
Assume that $\zeta_{k+1}^{},\zeta_{k+2}^{}, \cdots, \zeta_{k+i}^{}$ is a sample of $\zeta \stackrel[]{\text{i.i.d.}}{\sim} \mathrm{Uni}(0,1)$. Since $\e_{k+1\mid k}^{} = \beps_{k}^{}(i-1)$ when $\tau_{k}^{} = i-1, ~\forall i\in\{1, 2, \cdots, \mathrm{T}\}$, the stochastic triggering rule~\eqref{eqn:probabilistic_threshold_based_rule} can be rewritten as
\begin{align*}
	\delta_{k+i}^{} =
	\begin{cases}
		0 & \text{if} ~ \zeta_{k+i}^{} \leq e_{}^{-\lambda \langle \beps_{k+i-1}^{}(i-1), \beps_{k+i-1}^{}(i-1) \rangle} \;, \\
		1 & \text{otherwise} \;.
	\end{cases}
\end{align*}
For any given $\lambda > 0$, we compute:
\begin{align*}
	\mathbf{P}(\mathcal{E}_{i}^{}) =&\; \mathbf{P}( \delta_{k+1}^{}=0, \cdots, \delta_{k+i}^{}=0 ) \\
	=&\; \mathbf{P}\Bigg( \bigcap_{j=0}^{i-1} \zeta_{k+j+1}^{} \leq e_{}^{-\lambda \langle \beps_{k+j}^{}(j), \beps_{k+j}^{}(j) \rangle} \Bigg) \\
	=&\; \frac{\int_{\mathbb{R}_{}^{in}}^{}e_{}^{-\frac{1}{2}\chi_{}^{\top}(i-1)\big(2\lambda\I_{ni}^{} + \Sigma_{\epsilon}^{-1}(i-1)\big)\chi(i-1)}d\chi }{\sqrt{(2\pi)_{}^{in}\vert\Sigma_{\epsilon}^{}(i-1)\vert}}\\
	=&\; \frac{1}{\sqrt{\vert \I_{in}^{} + 2\lambda\Sigma_{\epsilon}^{}(i-1)\vert}} \;.
\end{align*}
This concludes the proof.
$\hfill\blacksquare$

{\it Proof of Lemma~\ref{lem:transition_probabilities}:}
We focus on the non-trivial cases where $i$ can take any value from $\{0,\cdots,\mathrm{T}-1\}$ and $j = 0$, as the remaining cases are evident from the structure of the Markov chain in Fig.~\ref{fig:Markov_Chain}. We first investigate the transition probability $p_{00}$. Since $\tau_{k}^{} = 0$ corresponds to $\delta_{k}^{} = 1$ as a consequence of~\eqref{eqn:timer}, we have:
\begin{align*}
p_{00} = &\; \mathbf{P}\big(\tau_{k+1}^{} = 0 \;\vert\; \tau_{k}^{} = 0 \big)  \\
		    = &\; \mathbf{P}\big( \delta_{k+1}^{} = 1 \;\vert\; \delta_{k}^{} = 1 \big)  \\
		  \stackrel{(a)}{=} &\; \mathbf{P}\big(  \delta_{k+1}^{} = 1 \big) = 1 - \mathbf{P}\big(  \delta_{k}^{} = 0 \big) \;,
\end{align*}
where ($a$) is true as $\delta_{k}$ is independent of the random variable $\boldeta_{k}^{}$. For any $i\in\{1,\cdots,\mathrm{T}-1\}$, we derive:
\begin{align*}
p_{i0} = &\; \mathbf{P}\big( \tau_{k+1}^{} = 0~\big\vert~\tau_{k}^{} = i \big)  \\
		  \stackrel{(b)}{=} &\; \mathbf{P}\big( \tau_{k+1}^{} = 0~\big\vert~\tau_{k}^{} = i, \cdots, \tau_{k-i+1}^{} = 1, \tau_{k-i}^{} = 0 \big)  \\
		  \stackrel{(c)}{=} &\; \mathbf{P}\big( \delta_{k+1}^{} = 1~\big\vert~\delta_{k}^{} = 0, \cdots, \delta_{k-i+1}^{} = 0, \delta_{k-i}^{} = 1 \big)  \\
		  \stackrel{(d)}{=} &\;  \mathbf{P}\big( \delta_{k+1}^{} = 1~\big\vert~\delta_{k}^{} = 0, \cdots, \delta_{k-i+1}^{} = 0 \big)  \\
		  = &\; \frac{\mathbf{P}\big( \delta_{k+1}^{} = 1, \delta_{k}^{} = 0, \cdots, \delta_{k-i+1}^{} = 0 \big)}{\mathbf{P}\big( \delta_{k}^{} = 0, \cdots, \delta_{k-i+1}^{} = 0 \big)} = 1 - \frac{\mathbf{P}\big( \mathcal{E}_{i+1}^{} \big)}{\mathbf{P}\big( \mathcal{E}_{i}^{} \big)} \;,
\end{align*}
where ($b$) comes from the Markov property, ($c$) is the result of~\eqref{eqn:timer}, and ($d$) holds since $\delta_{k-i}^{}$ is independent of the random variables $\boldeta_{k}^{}, \cdots, \boldeta_{k-i}^{}$. Using the result from Lemma~\ref{lem:probability}, we can straightforwardly compute the transition probabilities as given in the statement of the lemma.
$\hfill\blacksquare$

{\it Proof of Theorem~\ref{thm:communication_rate}:}
The proof of this theorem follows similar steps as in~\cite[pp.~98]{Dem:15}.
$\hfill\blacksquare$

{\it Proof of Theorem~\ref{thm:optimal_control_law}:}
Since~\eqref{eqn:probabilistic_threshold_based_rule} is a fixed scheduling rule with a predefined, constant parameter (i.e., $\lambda > 0$) and is a function of random variables $\{\x_{0}; \w_{0}^{}, \cdots, \w_{k}^{}; \v_{0}^{}, \cdots, \v_{k+1}^{}; \tilde{\x}_{0 \mid 0}^{s}, \cdots, \tilde{\x}_{k \mid k}^{s}; \tau_{0}^{}, \cdots, \tau_{k}^{}\}$, the transmission decisions $\sigma_{k}^{}$ (or consequently $\tau_{k}^{}$) are independent of the control law $\bu_{k}^{}$; see~\cite[Lemma~1]{MoH:13}. Therefore, the separation principle holds.

The proof of this theorem employs a dynamic programming argument; see~\cite{Ast:06}. Define the optimal value function $V_{k}^{}(\x_{k}^{})$ as follows:
\begin{multline}
	V_{k}^{}(\x_{k}^{}) = \min_{\bu_{k}^{}, \cdots, \bu_{N-1}^{}}^{} \\ \mathbf{E}\Bigg[ \x_{N}^{\top}Q_{f}^{}\x_{N}^{} + \sum_{t = k}^{N-1} \big( \x_{t}^{\top}Q\x_{t}^{} + \bu_{t}^{\top}R\bu_{t}^{} \big) \Bigg]. \label{eqn:cost_to_go}
\end{multline}
We claim that the solution of the functional equation~\eqref{eqn:cost_to_go} is a quadratic function of the form
\begin{align}
	V_{k}^{}(\x_{k}^{}) = \mathbf{E}\Big[ \x_{k}^{\top}S_{k}^{}\x_{k}^{} \mid \mathcal{I}_{k}^{c} \Big] + s_{k}^{} \;,
	\label{eqn:value_function}
\end{align}
where $S_{k}^{}$ is a non-negative definite matrix and $s_{k}^{}$ is a scalar. Indeed, this claim is clearly true for $k=N$ with the choice of parameters $S_{N}^{} = Q_{f}^{}$ and $s_{N}^{} = 0$. Suppose that the claim now holds for $k+1$. The value function at time-step $k$ is
\begin{align*}
&V_{k}^{}(\x_{k}^{}) =\; \min_{\bu_{k}^{}}\mathbf{E}\Big[ \x_{k}^{\top}Q\x_{k}^{} + \bu_{k}^{\top}R\bu_{k}^{} +  V_{k+1}^{}(\x_{k+1}^{}) \mid \mathcal{I}_{k}^{c} \Big] \\
&=\; \mathbf{E}\Big[ \x_{k}^{\top} \big( A_{}^{\top}S_{k+1}^{}A + Q - L_{k}^{\top}(B_{}^{\top}S_{k+1}^{}B + R)L_{k}^{} \big)\x_{k}^{} \mid \mathcal{I}_{k}^{c} \Big] \\
& + \mbox{Tr}\big( S_{k+1}^{}W \big) + s_{k+1}^{} + \mathbf{E}\Big[ \tilde{\x}_{k \mid k}^{s\top} L_{k}^{\top}(B_{}^{\intercal}S_{k+1}^{}B + R)L_{k}^{}\tilde{\x}_{k \mid k}^{s} \Big] \\
& + \mathbf{E}\Big[ \e_{k \mid k}^{\top} L_{k}^{\top}(B_{}^{\top}S_{k+1}^{}B + R)L_{k}^{}\e_{k \mid k}^{} \Big] \\
&\; + \min_{\bu_{k}^{}} \big( \bu_{k}^{} + L_{k}^{}\hat{\x}_{k \mid k}^{c} \big)_{}^{\top}(B_{}^{\top}S_{k+1}^{}B + R)\big( \bu_{k}^{} + L_{k}^{}\hat{\x}_{k \mid k}^{c} \big)
\end{align*}
%
which is obtained by writing $L_{k}^{} \triangleq (B_{}^{\intercal}S_{k+1}^{}B + R)_{}^{-1}B_{}^{\intercal}S_{k+1}^{}A \;$ and by replacing $\x_{k}^{}$ with $\e_{k \mid k}^{} = \x_{k}^{} - \hat{\x}_{k \mid k}^{s} - \tilde{\x}_{k \mid k}^{s}$. Hence, the minimum is obtained for
\begin{align*}
	\bu_{k}^{} = - L_{k}^{} \mathbf{E}\big[ \x_{k}^{} \mid \mathcal{I}_{k}^{c} \big] = - L_{k}^{} \hat{\x}_{k \mid k}^{c} \;.
\end{align*}
Consequently, the claim provided in~\eqref{eqn:value_function} is satisfied also for the time step $k$ for all $\x_{k}^{}$ if and only if
\begin{align*}
	S_{k}^{} =&\; A_{}^{\top}S_{k+1}^{}A + Q - L_{k}^{\top}(B_{}^{\top}S_{k+1}^{}B + R)L_{k}^{} \\
	s_{k}^{} =&\; s_{k+1}^{} + \mbox{Tr}\big( S_{k+1}^{}W \big) \\
	& \hspace{8mm} + \mbox{Tr}\Big( P_{k \mid k}^{s}L_{k}^{\top}(B_{}^{\top}S_{k+1}^{}B + R)L_{k}^{}\Big) \\
	& \hspace{8mm} + \mbox{Tr}\Big( L_{k}^{\top}(B_{}^{\top}S_{k+1}^{}B + R)L_{k}^{}\mathbf{E}\big[ \e_{k \mid k}^{}\e_{k \mid k}^{\top} \big] \Big)
\end{align*}
are satisfied. This concludes the proof.
$\hfill\blacksquare$

{\it Proof of Theorem~\ref{lem:Gaussianity_preservation}:}
The proof of this lemma follows similar arguments to~\cite[Lemma 4]{WRH+:16},  while also making use of the matrix inversion lemma.
$\hfill\blacksquare$

{\it Proof of Theorem~\ref{thm:control_performance}:}
The proof of ($a$) and ($b$) can be found in~\cite{Ast:06}. We, here, focus on only the proof of ($c$). Let us define $M_{k}^{}\triangleq L_{k}^{\top}(B_{}^{\top}S_{k+1}^{}B + R)L_{k}^{}$. As $N\rightarrow\infty$, similar to~\cite{Ast:06}, the expected minimum cost~\eqref{eqn:inf_expected_cost} can be written as
\begin{multline*}
	J_{\infty}^{} =\; \lim_{N\rightarrow\infty}^{}\frac{1}{N}J_{N}^{} =\; \mbox{Tr}(S_{\infty}^{}W) + \mbox{Tr}\Big( F_{\infty}^{s} M_{\infty}^{} \Big) \\ + \lim_{N\rightarrow\infty}^{}\frac{1}{N}\sum_{k=0}^{N-1} \mathbf{E}\Big[ \e_{k \mid k}^{\top} M_{k}^{}\e_{k \mid k}^{} \Big].
\end{multline*}
The last term in $J_{\infty}^{}$ ca be re-written as follows:
\begin{align}
	\lim_{N\rightarrow\infty}^{}& \mathbf{E}\bigg[\frac{1}{N}\sum_{k=0}^{N-1} \e_{k \mid k}^{\top} M_{k}^{}\e_{k \mid k}^{} \bigg] \nonumber\\
	=& \lim_{N\rightarrow\infty}^{} \mathbf{E}\bigg[\frac{1}{N}\sum_{k=0}^{N-1}\sum_{i=1}^{\mathrm{T}} \e_{k \mid k}^{\top} M_{k}^{}\e_{k \mid k}^{} \mathbbm{1}_{\{\tau_{k}^{}=i\}}^{} \bigg] \nonumber\\
	=& \sum_{i=1}^{\mathrm{T}}\bigg(\lim_{N\rightarrow\infty}^{} \frac{1}{N}\sum_{k=0}^{N-1} \mbox{Tr}\big( (M_{k}^{} - M_{\infty}^{})\Sigma_{e}^{}(i) \big) \mathbf{P}\big( \tau_{k}^{}=i \big) \nonumber\\
	& \hspace{5mm} + \lim_{N\rightarrow\infty}^{} \frac{1}{N}\sum_{k=0}^{N-1} \mbox{Tr}\big( M_{\infty}^{}\Sigma_{e}^{}(i) \big) \mathbf{P}\big( \tau_{k}^{}=i \big) \bigg). \label{eqn:eval_cost}
\end{align}
Since the pair $(A,B)$ is controllable and the pair $(A,Q_{}^{1/2})$ is observable, there exists a steady state $S_{\infty}^{}\in\mathbb{S}_{\succeq 0}^{n}$   for any initial matrix $S_{0}^{}\in\mathbb{S}_{\succeq 0}^{n}$. As a result, we have: $\lim\limits_{k\rightarrow\infty}^{}M_{k}^{} = M_{\infty}^{}$ (i.e., element-wise convergence). This implies that, for every $\varepsilon > 0$, there exists $N_{\varepsilon}^{}$ such that, for all $k > N_{\varepsilon}^{}$,
\begin{align*}
	\bigg\vert \sum_{s=1}^{n}\sum_{r=1}^{n} \big(m_{sr}^{k} - m_{sr}^{\infty}\big) \bigg\vert \leq \frac{\varepsilon}{2} \;,
\end{align*}
where $m_{sr}^k$ is the $(s,r)$-th entry of $M_k$ and $m_{sr}^\infty$ is the $(s,r)$-th entry of $M_\infty$.
The first term of~\eqref{eqn:eval_cost} can be upper-bounded as follows:
\begin{align*}
	&\frac{1}{N}\sum_{k=0}^{N-1} \mbox{Tr}\big( (M_{k}^{} - M_{\infty}^{})\Sigma_{e}^{}(i)\big) \mathbf{P}\big( \tau_{k}^{}=i \big) \\
	& \hspace{5mm}\leq \frac{1}{N}\Bigg\vert \sum_{k=0}^{N-1}\sum_{s=1}^{n}\sum_{r=1}^{n} \big(m_{sr}^{k} - m_{sr}^{\infty} \big) \Bigg\vert \max_{s,r\in\{1,\cdots,n\}}^{} \vert \sigma_{rs}^{}(i) \vert,
\end{align*}
where $\sigma_{rs}^{}(i)$ is the $(r,s)$-th entry of $\Sigma_{e}^{}(i)$. Let us define
\begin{align*}
	\tilde{m}_{k}^{} \triangleq \sum_{s=1}^{n}\sum_{r=1}^{n} m_{sr}^{k} \quad  \mbox{and} \quad \tilde{m}_{\infty}^{} \triangleq \sum_{s=1}^{n}\sum_{r=1}^{n} m_{sr}^{\infty}.
\end{align*}
%
%
Then, it is possible to divide the aforementioned sum (on the right hand side) into two parts:
\begin{align}
	\frac{1}{N}\Bigg\vert \sum_{k=0}^{N-1}\big( \tilde{m}_{k}^{} - \tilde{m}_{\infty}^{} \big)\Bigg\vert \leq &\; \frac{1}{N}\Bigg\vert \sum_{k=0}^{N_{\varepsilon}^{}-1}\big( \tilde{m}_{k}^{} - \tilde{m}_{\infty}^{} \big)\Bigg\vert \nonumber\\ 
	& + \frac{1}{N}\Bigg\vert \sum_{k=N_{\varepsilon}^{}}^{N-1}\big( \tilde{m}_{k}^{} - \tilde{m}_{\infty}^{} \big)\Bigg\vert. \label{eqn:cesaro_mean_inequality}
\end{align}
Let $\varepsilon > 0$. Choose $N_{\varepsilon}^{}$ large enough such that
\begin{align*}
	\frac{1}{N}\Bigg\vert\sum_{k=0}^{N_{\varepsilon}^{}-1}\big( \tilde{m}_{k}^{} - \tilde{m}_{\infty}^{} \big) \Bigg\vert &\leq \frac{1}{N}\sum_{k=0}^{N_{\varepsilon}^{}-1}\big\vert \tilde{m}_{k}^{} - \tilde{m}_{\infty}^{} \big\vert \\
		&\leq \max_{k\in\{0,\cdots,N_{\varepsilon}^{}-1\}}^{} \big\vert \tilde{m}_{k}^{} - \tilde{m}_{\infty}^{} \big\vert \frac{N_{\varepsilon}^{}}{N},
\end{align*}
holds for all $N > N_{\varepsilon}^{}$ and, if one chooses $N$ satisfying
\begin{align*}
	N > \tilde{N} \triangleq \frac{2N_{\varepsilon}^{}\max\limits_{k\in\{0,\cdots,N_{\varepsilon}^{}-1\}}^{} \big\vert \tilde{m}_{k}^{} - \tilde{m}_{\infty}^{} \big\vert}{\varepsilon} \;,
\end{align*}
then the first term of~\eqref{eqn:cesaro_mean_inequality} will be upper-bounded as follows:
\begin{align*}
	\frac{1}{N}\Bigg\vert \sum_{k=0}^{N_{\varepsilon}^{}-1}\big( \tilde{m}_{k}^{} - \tilde{m}_{\infty}^{} \big) \Bigg\vert \leq \frac{\varepsilon}{2} \;.
\end{align*}
We now bound the second term of~\eqref{eqn:cesaro_mean_inequality}:
\begin{align*}
	&\frac{1}{N}\Bigg\vert\sum_{k=N_{\varepsilon}^{}}^{N-1} \big( \tilde{m}_{k}^{} - \tilde{m}_{\infty}^{} \big) \Bigg\vert \leq\; \frac{1}{N}\sum_{k=N_{\varepsilon}^{}}^{N-1} \big\vert \tilde{m}_{k}^{} - \tilde{m}_{\infty}^{} \big\vert \\
	& \hspace{17mm} \leq\; \max_{k\in\{N_{\varepsilon}^{},\cdots,N-1\}}^{} \big\vert \tilde{m}_{k}^{} - \tilde{m}_{\infty}^{} \big\vert \frac{N - N_{\varepsilon}^{}}{N} \leq \frac{\varepsilon}{2}.
\end{align*}
%
%
It follows that, for all $N > \tilde{N}$, the inequality~\eqref{eqn:cesaro_mean_inequality} is bounded by $\varepsilon$ (i.e., an arbitrarily chosen upper-bound). In other words, the first term of~\eqref{eqn:eval_cost} converges to zero.

By the ergodic theorem~\cite[Theorem~5.3]{Bre:99} for Markov chains, we have:
\begin{align}
	\pi(i) = \lim_{N\rightarrow\infty}^{} \frac{1}{N}\sum_{k=0}^{N-1} \mathbbm{1}_{\{\tau_{k}^{}=i\}}^{},
\end{align}
which can be also represented, in the view of the bounded convergence theorem~\cite[pp. 138]{McW:13}, as
\begin{align}
	\pi(i) = \lim_{N\rightarrow\infty}^{} \frac{1}{N}\mathbf{E}\Bigg[\sum_{k=0}^{N-1} \mathbbm{1}_{\{\tau_{k}^{}=i\}}^{} \Bigg].
\end{align}
As a result, the second term of~\eqref{eqn:eval_cost} can be written as
\begin{align*}
	\lim_{N\rightarrow\infty}^{} \frac{1}{N}\sum_{k=0}^{N-1} &\mbox{Tr}\big( M_{\infty}^{}\Sigma_{e}^{}(i) \big) \mathbf{P}\big( \tau_{k}^{}=i \big) = \mbox{Tr} \big( M_{\infty}^{}\Sigma_{e}^{}(i) \big)\pi(i).
\end{align*}
%
%
This concludes our proof.
$\hfill\blacksquare$


\bibliographystyle{styles/IEEEtran}
\bibliography{references}

\end{document}